\documentclass[usenatbib,useAMS]{mn2e}
\usepackage{times}
\usepackage{epsfig}
\usepackage{amsmath}
\usepackage[usenames,dvips]{color}
\usepackage{subfigure}
\usepackage{rotating}
\usepackage{graphicx}
\usepackage{lscape}
\usepackage{pdflscape}
\graphicspath{{fig/}}

\begin{document}
\title[Detecting stellar spots through polarimetry observations of microlensing events in caustic-crossing]{Detecting stellar spots through polarimetry observations of microlensing events in caustic-crossing}
\author[Sedighe Sajadian]{Sedighe Sajadian$^{1,2}$\thanks{E-mail:
sajadian@ipm.ir}\\
$^{1}$School of Astronomy, IPM (Institute for Research in
Fundamental Sciences), P.O. Box 19395-5531, Tehran, Iran\\
$^{2}$School of Physics, Damghan University, P.O. Box 41167-36716,
Damghan, Iran}

\maketitle
\begin{abstract}
In this work, we investigate if gravitational microlensing can
magnify the polarization signal of a stellar spot and make it be
observable. A stellar spot on a source star of microlensing makes
polarization signal through two channels of Zeeman effect and
breaking circular symmetry of the source surface brightness due to
its temperature contrast. We first explore the characteristics of
perturbations in polarimetric microlensing during caustic-crossing
of a binary lensing as follows: (a) The cooler spots over the
Galactic bulge sources have the smaller contributions in the total
flux, although they have stronger magnetic fields. (b) The maximum
deviation in the polarimetry curve due to the spot happens when the
spot is located near the source edge and the source spot is first
entering the caustic whereas the maximum photometric deviation
occurs for the spots located at the source center. (c) There is a
(partial) degeneracy for indicating spot's size, its temperature
contrast and its magnetic induction from the deviations in light or
polarimetric curves. (d) If the time when the photometric deviation
due to spot becomes zero (between positive and negative deviations)
is inferred from microlensing light curves, we can indicate the
magnification factor of the spot, characterizing the spot properties
except its temperature contrast. The stellar spots alter the
polarization degree as well as strongly change its orientation which
gives some information about the spot position. Although, the
photometry observations are more efficient in detecting stellar
spots than the polarimetry ones, but polarimetry observations can
specify the magnetic field of the source spots.
\end{abstract}

\section{Introduction}
Gravitational microlensing is an astronomical phenomenon in which
the light of a background star is magnified due to passing through
the gravitational field of a foreground object by producing
distorted images \cite{Einstein36}. This phenomenon has many
applications in detecting extra-solar planets, measuring the mass of
isolated stars, studying the stellar atmospheres, etc. (see, e.g.
Mao 2012,Gaudi 2012).

One of features of gravitational microlensing is that it causes a
net polarization for source stars
\cite{schneider87,simmons95a,Bogdanov96}. There is a local
polarization over the star surface due to the electron scattering in
its atmosphere \cite{chandrasekhar60}. Since, these local
polarizations have symmetric orientations with respect to the source
center, the net polarization of a distant star is zero. Some effects
can break this symmetry and create a net polarization for distant
stars such as circum-stellar disk of dust grains around the central
star \cite{Drissen1989,Akitaya2009}, stellar spots, magnetic field
and lensing effect. During a microlensing event, the circular
symmetry of source star is broken and result is a net polarization
for the source star. Measuring polarization during microlensing
events helps us to evaluate the finite source effect, the Einstein
radius and the limb-darkening parameters of source star
\cite{yoshida06,Agol96,schneider87}.

In single microlensing events, the polarimetric signal reaches to
its maximum amount, i.e. about one percent, when the impact
parameter of the source center i.e. $u_{cm}$ reaches to $\sim 0.96
\rho_{\star}$ where $\rho_{\star}$ is the projected source radius,
normalized to the Einstein radius \cite{schneider87}. But, the
probability of happening a microlensing event with an impact
parameter as smaller as $u_{cm}<\rho_{\star}$ is low. Whereas, in
binary microlensing events whenever the source star crosses the
caustic curve the polarization signal as high as one percent can be
obtained \cite{Agol96}. Indeed, the point-like caustic in a single
lens converts to some closed curves in binary lensing which is
extended about the angular Einstein radius (see e.g. Schneider and
Weiss 1986). 

Consequently, whenever the source crosses the caustic curves, due to
the high gradient of magnification and polarization, the anomalies
over the source surface can be resolved through photometric or
polarimetric observations. One of the second-order perturbations
over the source surface is the stellar spot. Indeed, a magnetic
field over the source surface generates convective motions which
make a temperature contrast with the source surface temperature,
so-called as a stellar spot (e.g. Berdyugina 2005). A stellar spot
produces a polarization signal through two channels (i) Zeeman
effect due to its magnetic field and (ii) circular symmetry breaking
of the source surface brightness due to the temperature contrast
with the host star. Comparing these two effects, the first one has a
dominant contribution in polarization signal. A stellar spot over
the source surface disturbs the light and polarimetric curves of
microlensing events. Detecting and characterizing stellar spots
through their effects on the photometric light curves during
microlenising events due to a point-like lens and binary lenses were
extensively studied in (Heyrovsk\'y \& Sasselov (2000),Han et al.
(2000), Chang \& Han (2002), Hendry et al. (2002), Hwang \& Han
(2010)). Here, we investigate detecting and characterizing stellar
spots over the source surface through their effects on the
\emph{polarimetric curves} in microlensing events.

Although, most spot-induced polarization signals from the Galactic
bulge sources are too weak to be detected by themselves, the lensing
effect can magnify these signals and makes them be observable.
Studying spot-induced perturbations in polarimetric microlensing due
to a point-like lens was first investigated by Sajadian and Rahvar
(2014). They also noticed that there is an orthogonal relation
between the polarization orientation and the astrometric centroid
shift vector in microlensing events due to a point-like lens and
binary lenses except near the fold of caustic and investigated the
advantages of this orthogonality for detecting the anomalies of the
source surface. Here, we study detecting and characterizing the
stellar spot through polarimetric microlensing in caustic-crossing.
We first explore the characteristics of spot-induced perturbations
on the polarization degree and its orientation in binary
microlensing events. Then, by performing a Monte Carlo simulation
from binary microlensing events with spotted sources during
caustic-crossing we estimate the probability of detecting a stellar
spot through polarimetric microlensing during caustic-crossing and
the number of events.

In the following section, we study the stellar spot effects on (i)
magnification factor, (ii) polarization degree and (iii)
polarization orientations. In section (\ref{mont}) the detectability
of the polarimetric and photometric deviations due to the stellar
spots in microlensing events during caustic-crossing is studied by
performing a Monte Carlo simulation. We explain the results in the
last section.

\section{spot-induced perturbation in polarimetric microlensing}\label{two}
In order to describe the polarized light we use the Stokes
parameters of $S_{I}$, $S_{Q}$, $S_{U}$ and $S_{V}$. These
parameters describe the total intensity, two components of linear
polarization and circular polarization over the source surface,
respectively \cite{Tinbergen96}. The polarization degree $(P)$ and
the polarization angles $\phi_{p}$ and $\theta_{p}$ as functions of
Stokes parameters are given by \cite{chandrasekhar60}:
\begin{eqnarray}
P&=&\frac{\sqrt{S_{Q}^{2}+S_{U}^{2}+S_{V}^{2}}}{S_{I}},\nonumber\\
\phi_{p}&=&\frac{1}{2}\tan^{-1}{\frac{S_{U}}{S_{Q}}},\nonumber\\
\theta_{p}&=&\frac{1}{2}\tan^{-1}{\frac{S_{V}}{\sqrt{S_{Q}^2+S_{U}^2}}}.
\end{eqnarray}
By considering the linear polarization of light scattering in a
stellar atmosphere, the circular polarization is zero $S_{V}=0$. In
microlensing events, the Stokes parameters during the magnification
are given by:
\begin{eqnarray}\label{tsparam}
\left( \begin{array}{c}
S_{Q,\star}\\
S_{U,\star}\end{array}\right)&=&\rho^2_{\star}\int_{0}^1\rho~d\rho\int_{-\pi}^{\pi}d\phi I_{-}(\mu) A(u) \left( \begin{array}{c} -\cos 2\phi \\
\sin 2\phi \end{array} \right),\nonumber\\
S_{I,\star}&=&~\rho^2_{\star}\int_{0}^{1}\rho
d\rho\int_{-\pi}^{\pi}d\phi I(\mu)~ A(u),
\end{eqnarray}
where $\rho_{\star}$ is the projected source radius which is
normalized to the Einstein radius, $\rho$ is the distance from the
centre of the stellar disk to each element over the source surface
which is normalized to the source radius, $\mu=\sqrt{1- \rho^{2}}$,
$\phi$ is the azimuthal angle between the lens-source connection
line and the line from the center of coordinate to each element over
the source surface, $u=(u_{cm}^2+ \rho^2 \rho^{2}_{\star}+2 \rho
\rho_{\star} u_{cm} \cos\phi)^{1/2}$ is the distance of each
projected element over the source surface with respect to the lens
position, $u_{cm}$ is the impact parameter of the source center and
magnification factor for simple microlensing is $A(u)=\frac{u^2+2}{u
\sqrt{u^2+4}}$. The amounts of $I(\mu)$ and $I_{-}(\mu)$, as the
total and polarized light intensities, by assuming the electron
scattering in spherically isotropic scattering atmosphere of an
early-type star on the emitted light, were evaluated by
Chandrasekhar (1960) as follows:
\begin{eqnarray}
I(\mu)&=&I_{\star} (1-c_{1}(1-\mu) ),\nonumber\\
I_{-}(\mu)&=&I_{\star}c_{2}(1-\mu),
\end{eqnarray}
where $I_{\star}$ is the source intensity at the disk center,
$c_{1}=0.64$ and $c_{2}=0.032$ \cite{schneider87}. Here, we focus on
the giant sources with finite size effect for which the
limb-darkening coefficients are different a bit. For these sources,
we set $c_{1}=0.7$ \cite{zub2009} and $c_{2}=0.04$.

\begin{figure}
\begin{center}
\psfig{file=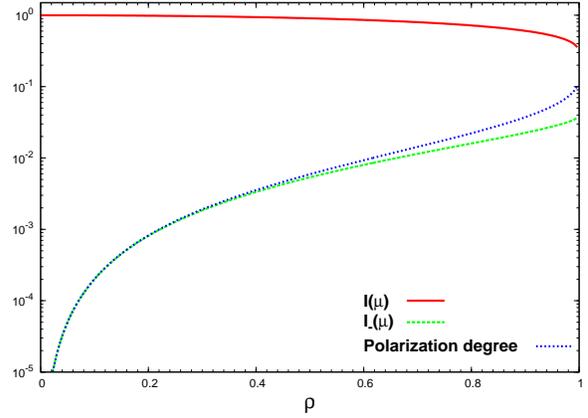,angle=270,width=8.cm,clip=0}\caption{$I(\mu)$
the total intensity (solid red line), $I_{-}$ the polarized
intensity (green dashed line) normalized to the source intensity at
the disk center, i.e. $I_{\star}$, and the polarization degree (blue
dotted line) versus the distance from the source center normalized
to the source radius i.e. $\rho$.}\label{fig1}
\end{center}
\end{figure}

Figure (\ref{fig1}) represents $I(\mu)$ (red solid line),
$I_{-}(\mu)$ (green dashed line), normalized to the intensity of the
source at the disk center, versus the distance from the source
center normalized to the source radius i.e. $\rho$. According to
this figure, considering the limb-darkening effect the total
intensity $I(\mu)$ decreases a bit whereas the polarized intensity
$I_{-}(\mu)$ increases from the center to the limb. If there is no
lensing effect i.e. $A(u)=1$, the local polarization degree is equal
to $P=I_{-}(\mu)/I(\mu)$. Here, over the star surface the
polarization degrees enhance from the center to limb as shown in
Figure (\ref{fig1}) with blue dotted line. Also, the polarization
vector in each point of the source surface is tangent to a circle
whose center coincides to the source center and passes from that
point. The overall polarization signal of an ideal source is zero,
because of the symmetric orientations of local polarizations with
respect to the source center.

Now, let us consider a stellar spot on the source surface. To
specify the spot in our simulation, we use four parameters for spots
as: (i) the size of spot, (ii) the location of spot on the source
star, (iii) its magnetic field and (iv) the temperature contrast of
the spot with respect to the host star. These parameters are not
completely independent and  there is a correlation relation between
the magnetic field of the spot and its temperature contrast with
respect to the host star \cite{starspot}. Also, we consider the
following limitations for modeling the spot polarization: (i) There
is one circular spot on the source surface, (ii) the magnetic field
over the spot has a constant amount and its direction is radial,
(iii) the magnetic field is high enough to ignore the Hanle effect
(see e.g. Stenflo (2013)), (iii) we take the Minle-Eddington model
for the stellar atmosphere in which the line source function is
linear versus the optical depth, (iv) the ratio of the line and
continuous absorption coefficients is constant, (v) the intensity
over the spot is constant. Also we assume a black body radiation for
the spot and its host star to indicate their intensities and finally
(vi) we estimate the broad band Stokes intensities by integrating
over the wavelength pass band, taking into account the Stokes
intensities in a spectral line.

The radius of the spot is $r_{s}$ and the angular radius of spot in
coordinate located at the centre of source star is given by
$\theta_{0}=\sin^{-1} (r_{s}/R_{\star})$. We preform a coordinate
transformation to obtain the position of spot on the lens coordinate
set two axes of which describe the sky plane. In this regard, we use
the two rotation angles of $\theta_{s}$ around $y$-axis and
$\phi_{s}$ around $z$-axis. The more details about modeling of the
spot in our simulation can be found in Sajadian and Rahvar (2014).

To calculate the modified Stokes parameters, we integrate over the
source surface from the Stokes intensities. Since, the Stokes
intensities over a given spot are different from those over the
other points of the source surface, we take into account the
combination of the terms from spot and source star, using a step
function. The overall Stokes parameters $S'_{Q}$, $S'_{U}$, $S'_{V}$
and $S'_{I}$ are given by:
\begin{eqnarray}\label{spott}
\left(\begin{array}{c}S'_{Q}\\S'_{U}\\S'_{V}\\S'_{I}\end{array}\right)=
\rho_{\star}^{2}~\int_{0}^{1}\rho d\rho~\int_{-\pi}^{\pi} d\phi
A(u)\{~~~~~~~~~~~~~~~~~~~~~~~~~~~~~~~~~~~~~~~~~~~\nonumber\\(1-\Theta(\rho,\phi))
\left(\begin{array}{c}-I_{-}(\mu)\cos2\phi\\I_{-}(\mu)\sin2\phi\\
0\\ I(\mu)\end{array}\right)+\Theta(\rho,\phi)~\left(\begin{array}{c}I_{Q,s}\\I_{U,s}\\
I_{V,s}\\I_{I,s}\end{array}\right)~\},\end{eqnarray} where
$\Theta(\rho,\phi)$ is that step function which is equal to one
within the spot area and zero for the other points. $I_{Q,s}$,
$I_{U,s}$, $I_{V,s}$ and $I_{I,s}$ are the Stokes intensities over
the spot which are measured according to the spot position, it
magnetic field and its intensity. Indeed, the magnetic field through
the Zeeman effect on the spectral lines causes a net polarization
for the spot which is mostly circular (see e.g. Illing et al. 1975).
The more details about calculating these Stokes intensities by
considering the mentioned limitations over the source and its
atmosphere are brought in Appendix section.

The spot-induced perturbations on the magnification factor,
polarization degree and polarization angles are explained in the
following subsections respectively.

\subsection{Perturbation on the magnification factor}
The spot-induced perturbation on microlensing light curve due to a
point-like lens was first investigated by Heyrovsk\'y and Sasselov
(2000). They concluded that the photometric spot effect can
numerically reach the fractional spot radius. Then, the feasibility
of spot detection, studying the spot-induced perturbations in binary
microlensing events during the caustic crossing and delectability of
spot in different wavelengths have been discussed in details by a
number of authors \cite{Han2000,Chang2002,Han2010,Hendry2002}.

To calculate the magnification factor of a spotted source, we assume
that the intensity of the spot, i.e. $I_{I,s}$, is almost constant
over the spot area. Hence, we can re-write the fourth modified
Stokes parameter $S'_{I}$ (in equation \ref{spott}), by factoring
out $I_{I,s}$ as:
\begin{eqnarray}
S'_{I}&\cong&
S_{I,\star}-b~I(\mu_{s})~\rho_{\star}^{2}\int_{0}^{1}~\rho~d\rho~\int_{-\pi}^{\pi}
d\phi~A(u)~\Theta(\rho,\phi)\nonumber\\&=&S_{I,\star}-b~S_{I,s},
\end{eqnarray}
where we define $b~\equiv~(I(\mu_{s})-I_{I,s})/I(\mu_{s})$,
$I(\mu_{s})$ is the source intensity at the geometrical location of
the spot center i.e. $(\theta_{s},\phi_{s})$, $\mu_{s}=\cos
\theta_{s}$, $S_{I,\star}$ is the Stokes parameter due to the source
without any spot, given by equation (\ref{tsparam}) and $S_{I,s}$ is
the Stokes parameter of the spot itself considering $I(\mu_{s})$ as
its intensity. Integrating is done over the source area. In that
case, the magnification factor of a spotted source is given by:
\begin{eqnarray}
A'=\frac{S_{I,\star}-b~S_{I,s}}{S_{I,\star,0}-b~S_{I,s,0}}=A\frac{1-b\beta}{1-b\delta},
\end{eqnarray}
where the index $0$ of the Stokes parameters refers to the ones
without lensing effect. $A=S_{I,\star}/S_{I,\star,0}$ is the
magnification factor of the source without any spot,
$\beta=S_{I,s}/S_{I,\star}$ and $\delta=S_{I,s,0}/S_{I,\star,0}$.
The magnification factor $A'$ up to the third order in $\delta$ in
terms of un-perturbed magnification factor $A$ and perturbation
terms is given by:
\begin{eqnarray}\label{eq7}
A'=A[1+ b(\delta-\beta)+b^{2}\delta(\delta-\beta)+
b^{3}\delta^{2}(\delta-\beta) + .... ].
\end{eqnarray}
Here, a spot on the surface of source star has two effects: (i)
decreasing the baseline flux of the source which enhances the
magnification factor by $b\delta+b^2\delta^2+...$ and (ii)
decreasing the magnified light which decreases the magnification
factor about $-b\beta-b^2\beta\delta- ...$. Before the spot reaches
to the caustic curve, the magnification factor enhances due to the
first effect. When the spot light is being magnified due to passing
over the caustic line or approaching to the lens position, the
second effect dominates and the total magnification factor
decreases. When these two effects have the same amounts i.e.
$\beta=\delta$, the perturbation terms vanish. In this time,
$A'=A=A_{s}$ where $A_{s}=\frac{S_{I,s}}{S_{I,s,0}}$ which means
that the magnification factor of the source is equal to the
magnification factor of the spot itself. This time does not depend
on the relative difference in the intensity of source and spot i.e.
$b$ and just depends on the spot area and its location with respect
to the lens or caustic line. If the source radius and the lens
position (or the caustic configuration in binary lensing) in this
time are deduced from the photometric observations, we can assign
the magnification factor of the spot itself which helps us to
characterize the spot properties.

\begin{figure*}
\begin{center}
\subfigure[] { \label{fig2a}
\psfig{file=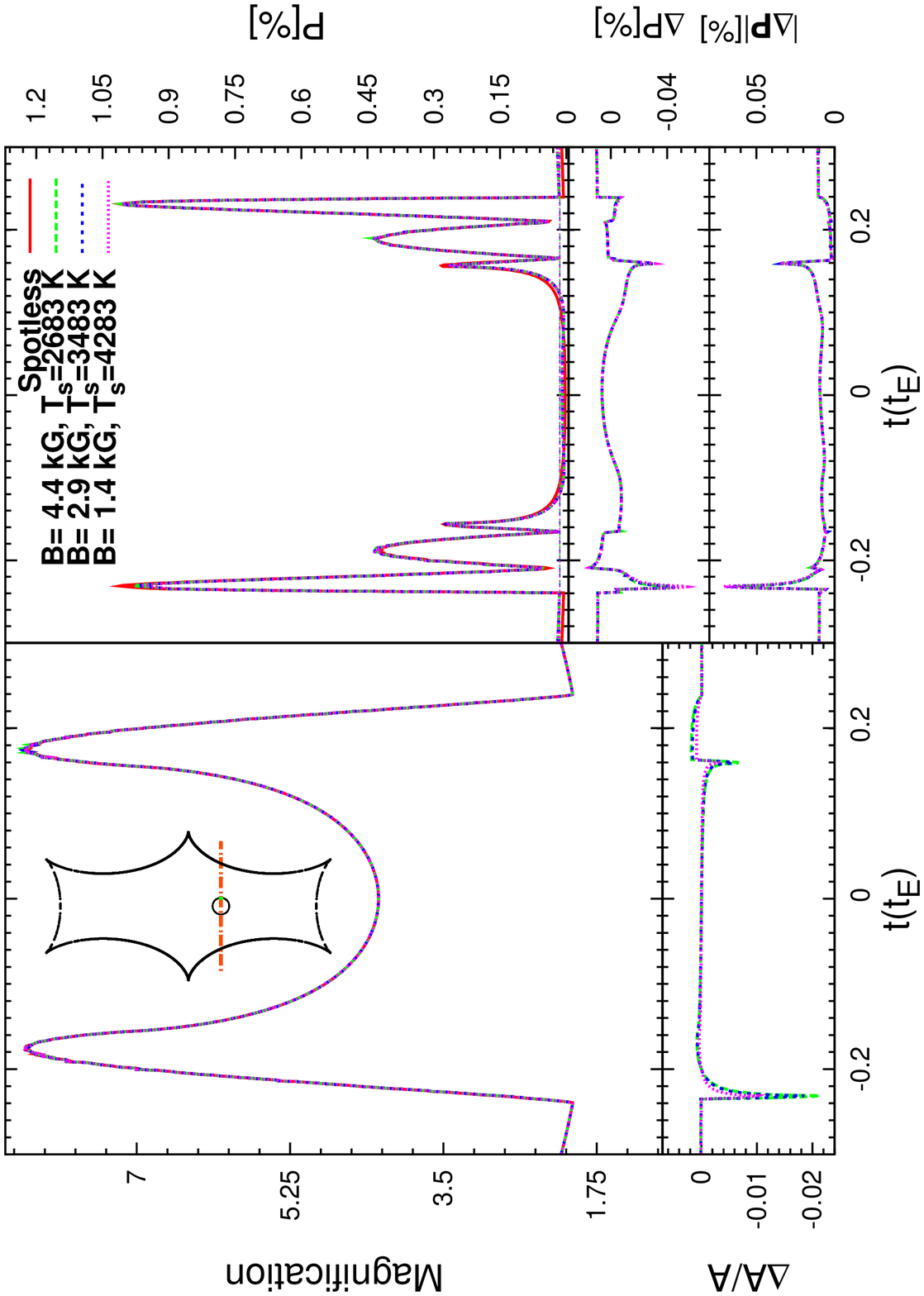,angle=0,width=0.41\textwidth,clip=0}}
\subfigure[] { \label{fig2b}
\psfig{file=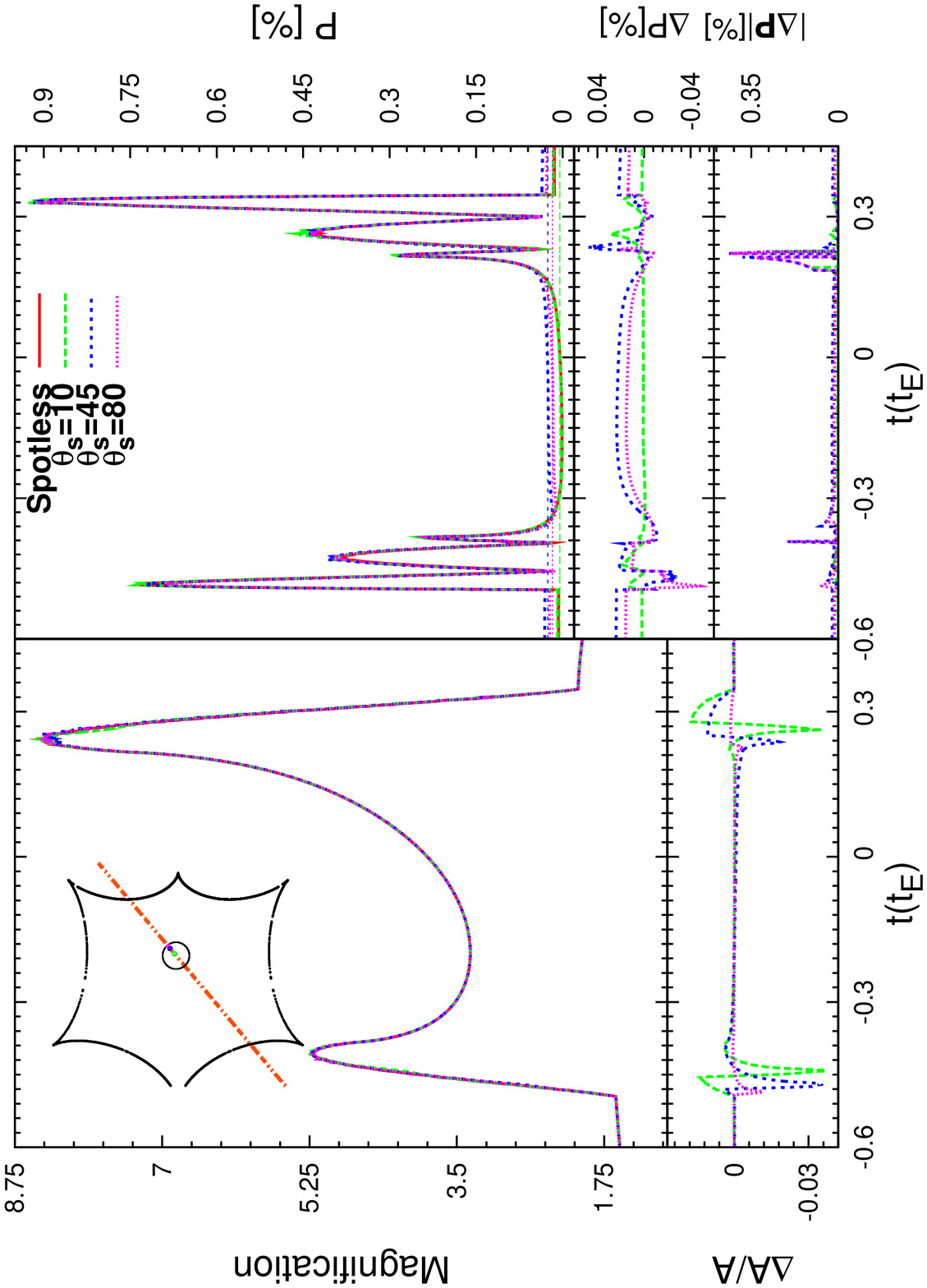,angle=0,width=0.41\textwidth,clip=0}}
\subfigure[] { \label{fig2c}
\psfig{file=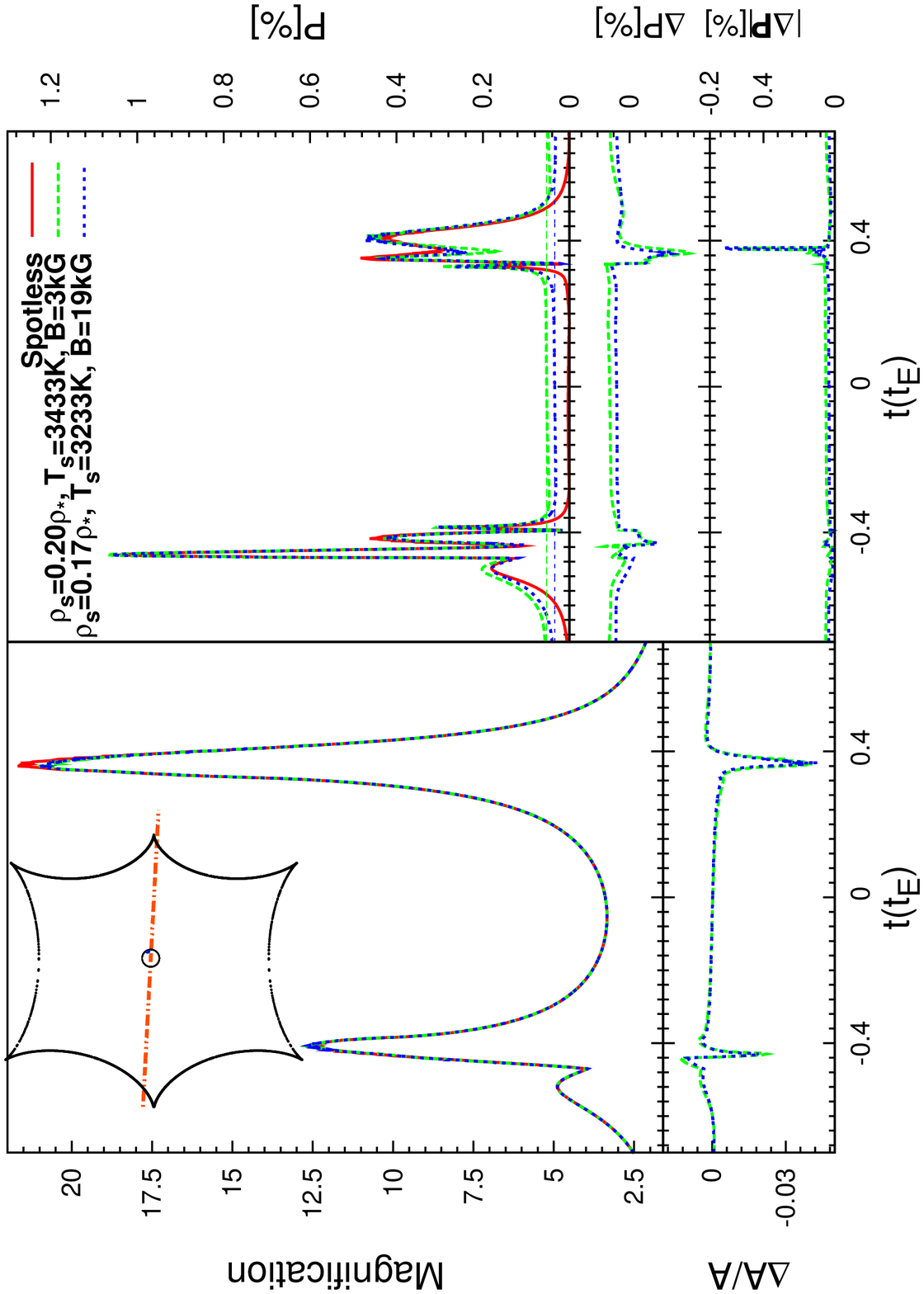,angle=0,width=0.41\textwidth,clip=0}}
\subfigure[] { \label{fig2d}
\psfig{file=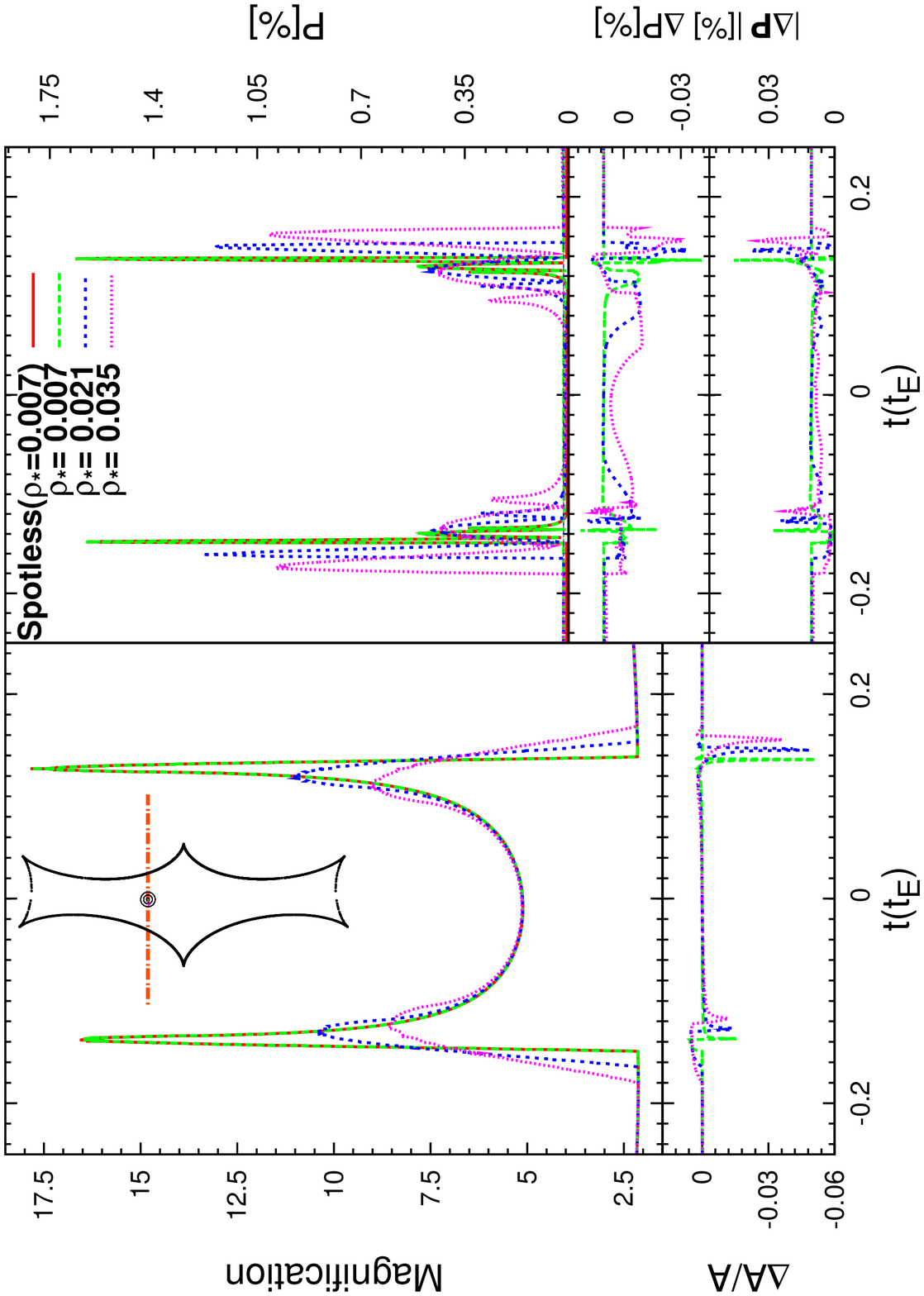,angle=0,width=0.41\textwidth,clip=0}}
\caption{Example microlensing events affected with source spot. In
every subfigure, the light curves and polarimetric curves are shown
in left and right panels. The source (black circle) and its spot
located over the source surface, caustic curve (black curve) and
source centre trajectory projected in the lens plane (red
dash-dotted line) are shown with insets in the left-hand panels. The
thinner lines in right panel show the spot polarization signal
without lensing effect. The simple models without spot effect are
shown by red solid lines. The photometric and polarimetric residuals
with respect to the simple models are plotted in bottom panels. Note
that the top polarimetric residuals are the residual in the
polarization degree $\Delta P$ and the bottom ones are the absolute
value of the residual in the polarization vector
$|\boldsymbol{\Delta P}|$. The parameters used to make these
microlensing events can be found in Table (1).} \label{fig2}
\end{center}
\end{figure*}

\subsection{Perturbation on the polarization degree}
Here, we investigate the spot-induced perturbations on polarization
degree during a microlensing event, some prospects are explained in
the following.

(a) Berdyugina (2005) noticed that there are some correlation
relations between the spot parameters such as its magnetic field,
temperature contrast and the source photosphere temperature. She
plotted the data of magnetic field measurement $B(kG)$ and the spot
temperature contrast $\Delta T(K)$ versus the photosphere
temperature $T_{\star}(K)$ for some active dwarfs and giants and
concluded that the hotter giants have the weaker magnetic field
whereas their spots have the higher temperature contrast. We can fit
two useful functions to her data which are given by:
\begin{eqnarray}\label{BT}
\Delta T(K)&=&1.26\times 10^{-4} T_{\star}^{2}-0.53
T_{\star}+787.40,\nonumber\\
B(kG)&=&-1.06\times 10^{-3}~T_{\star}+7.46.
\end{eqnarray}
According to these relations, for a source star with a constant
photosphere temperature, the cooler stellar spots have the stronger
magnetic fields and as a result the higher polarization signals.

On the other hand, the Stokes intensities of the spot, i.e.
$I_{i,s}$ (for $i=Q,U,V,I$), are proportional to the continuum,
local intensity of the spot, i.e. $I_{c}$, (for more details see
Appendix section). But, in the microlensing observations from the
sources in the Galactic bulge, we receive the total flux of the
source and its spot. Hence, the darker spots have the smaller
contributions to the total flux whereas they intrinsically have
larger polarization signals due to the stronger magnetic fields.
Here, we investigate which of these two effects, i.e. (1) the spot
magnetic field and (2) the spot flux contrast, has more contribution
in the polarimetric perturbations in microlensing events.
\begin{table*}
\begin{center}
\begin{tabular}{|c|c|c|c|c|c|c|c|c|c|c|c|c|c|c|c|}
& Figure number & $M_{\star}(M_{\odot})$ & $R_{\star}(R_{\odot})$ &
$T_{\star}(K)$ & $\rho_{\star}$ & $r_{s}(R_{\star})$ &
$T_{s}(K)$ & $B(kG)$ & $\theta_{s}^{\circ}$ & $\phi_{s}^{\circ}$& $u_{cm}$  &$\xi^{\circ}$ &  $d$ & $q$ &\\
\hline\hline
& $2(a)$ & $2.6$ & $24.9$ & $4314$  & $0.04$ & $0.35$ & $--$ & $--$ & $70$ & $0$ & $-0.15$ & $0$ & $1$ & $1$ & \\
\hline
& $2(b)$ & $2.8$ & $28.4$ & $4302$ & $0.05$ & $0.35$ & $3477$ & $2.9$ & $--$ & $40$ & $0.01$ & $40$ & $1.2$ & $0.5$ &  \\
\hline
& $2(c)$ & $1.9$ & $17.2$ & $4206$  & $0.03$ & $0.30$ & $--$ & $--$ & $55$ & $23$ & $0.01$ & $-3$ & $1.2$ & $0.8$ & \\
\hline
& $2(d)$ & $1.2$ & $4.5$ & $4492$  & $--$ & $0.3$ & $3495$ & $2.7$ & $-45$ & $10$ & $0.15$ & $0$ & $0.9$ & $0.9$ & \\
\hline
& $(3)$ & $2.8$ & $28.4$ & $4302$ & $0.05$ & $0.35$ & $3477$ & $2.9$ & $40$ & $0$ & $0.01$ & $40$ & $1.2$ & $0.5$ &  \\
\hline
\end{tabular}
\end{center}
\caption{The table contains the parameters used to make the
microlensing events shown in Figures \ref{fig2a}, \ref{fig2b},
\ref{fig2c}, \ref{fig2d} and (\ref{fig3}) respectively. These
parameters are the source mass $M_{\star}(M_{\odot})$, the source
radius $R_{\star}(R_{\odot})$, the photosphere temperature of the
source $T_{\star}(K)$, the projected source radius in the lens
plane, normalized to the Einstein radius $\rho_{\star}$, the spot
radius normalized to the source radius $r_{s}(R_{\star})$, the spot
temperature $T_{s}(K)$, the spot magnetic field $B(kG)$, the spot
projection angles $\theta_{s}^{\circ}$ and $\phi_{s}^{\circ}$, the
impact parameter of the source center $u_{cm}$, the angle between
the trajectory of source star and the binary axis $\xi^{\circ}$, the
distance between two lens normalized to the Einstein radius $d$ and
the mass ratio of two lens $q$ respectively. The giant stars as
sources are chosen from a synthetic distribution for giant stars
made from Besan\c{c}on model (Robin et al. 2003). For these figures,
we also set the mass of the primary lens $M_{l}=0.3~M_{\odot}$, the
lens and source distance from the observer $D_{l}=6.5~kpc$ and
$D_{s}=8.0~kpc$ and the limb-darkening coefficients $c_{1}=0.7$ and
$c_{2}=0.04$.}\label{tab1}
\end{table*}

For this aim, in Figure \ref{fig2a} we show a polarimetric
microlensing event affected by the source spot. We consider a giant
star as the source with the parameters $T_{\star}(K)=4314$,
$M_{\star}(M_{\odot})=2.6$ and $R_{\star}(R_{\odot})=24.9$. Also we
assume that the source has a spot with the parameters $\Delta
T(K)=831$ and $B(kG)=2.9$, chosen according to the equations
(\ref{BT}). In addition, we consider a cooler spot with a stronger
magnetic field and the hotter spot with a weaker magnetic field. The
light curves and polarimetric curves for these three cases are shown
in left and right panels of this figure. The source (black circle)
and its spot located over the source surface, caustic curve (black
curve) and source center trajectory projected in the lens plane (red
dash-dotted line) are shown with an inset in the left-hand panel.
The thinner lines in right panel show the intrinsic polarization
signals of the spotted source. The simple models without spot effect
are shown by red solid lines. The photometric and polarimetric
residuals with respect to the simple models are plotted in bottom
panels. Note that the polarimetric residuals are the residual in the
polarization degree without considering the polarization angles i.e.
$\Delta P$ and the absolute value of the residual in the
polarization vector i.e. $|\boldsymbol{\Delta P}|$ whose definition
is brought in equation (\ref{delp}). The parameters used to make
this microlensing event can be found in Table (\ref{tab1}). To
calculate the magnification factor of each element over the source
star, we use the adaptive contouring algorithm \cite{Dominik2007}.
According to this figure three spots with different magnetic fields
and temperature contrasts have almost the same intrinsic
polarization signals and as a result the same polarimetric
deviations in the polarimetric microlensing. Hence, these two
effects whereas act reversely have the same strengths. The
photometric signature of the spot depends on the temperature
contrast and enhances with increasing it.

The polarization signal of spot maximizes when the spot is crossing
the caustic curve. If the spot is located on the source edge and the
source is entering into the caustic from that edge (near the spot),
the spot-induced perturbation is so higher than when that source is
coming out from that edge. Because, in the first case the light from
the other points over the source surface is not magnified.
Accordingly, in the first caustic-crossing of Figure \ref{fig2a},
the spot-induced perturbation is higher than that in the second one.

Also, the duration of the polarimetric signal depends on the
location of spot with respect to the caustic line, the size of spot
and its host star and it does not depend on the temperature contrast
and the magnetic field.

(b) The effect of the projection angle $\theta_{s}$ on the
spot-induced polarimetric deviation: the amount of the circular
Stokes intensity in a spectral line due to Zeeman effect is
proportional to $\cos \theta_{s}$ and as a result increases by
decreasing $\theta_{s}$, whereas the linear Stokes intensities in a
spectral line are proportional to $\sin^{2} \theta_{s}$ and decrease
by decreasing $\theta_{s}$ (see equations \ref{A7} and \ref{A5} in
Appendix section). Hence, the polarization signals due to face-on
spots are more circular and those due to spots with larger
projection angles are more linear. As mentioned in the Appendix
section, the broadband circular Stokes intensity is so smaller than
that in a spectral line by one order of magnitude whereas the linear
ones have almost the same amounts as those in a spectral line.
Because, the circular polarization in a spectral line has an
anti-symmetric shape with respect to the wavelength. Consequently,
stellar spots with the smaller projection angle $\theta_{s}$ have
lower polarized Stokes intensities.

In Figure \ref{fig2b} a polarimetric microlensing event of a spotted
source is represented with three different amounts of the projection
angle $\theta_{s}$. According to this figure, the intrinsic
polarization signal due to the spot without lensing for
$\theta_{s}=10^{\circ}$ (about $0.005$ per cent) is so smaller than
that with $\theta_{s}=45^{\circ}$ (about $0.026$ per cent), shown
with thin, straight lines. On the other hand, increasing the
projection angle $\theta_{s}$ decreases the spot area and the total
intensity considering limb-darkening effect and as a result
increases $S'_{I}$ (which decreases the polarization signal).
Therefore, by increasing the projection angle, the polarization
signal due to stellar spots without lensing increases and then
decreases. The polarization signal without lensing due to the spot
with $\theta_{s}=80^{\circ}$ is equal to $0.018$ per cent, lower
than that with $\theta_{s}=45^{\circ}$.

However, the deviations in the polarimetric microlensing due to the
spot with $\theta_{s}=80^{\circ}$ is a bit larger than that due to
the spot with $\theta_{s}=45^{\circ}$ which is due to the finite
size effect. The larger projection angle, the smaller area of the
spot. In microlensing events, finite size effect always shrinks
sharp signals.

In the photometric curves, increasing the projection angle
$\theta_{s}$ decreases the spot-induced photometric signal. Because,
the spot area decreases by increasing the projection angle by $\cos
\theta_{s}$ which decreases the spot-induced perturbation signal.
However, the total intensity decreases from the center to the limb
of the source star due to limb-darkening effect (see Figure
\ref{fig1}), but this effect is so smaller than the first one. If
the area of the spot is fixed by considering the limb-darkening
effect, the photometric signal due to the spot does not change
significantly (see e.g. Heyrovsk\'y \& Sasselov (2000)). Therefore,
the spots located at the source center make higher photometric
signals as shown in Figure \ref{fig2b}. This points was also noticed
by Hendry et al. (2002).

(c) There is a partial degeneracy for indicating the spot size and
its temperature difference with respect to the host star from the
photometric deviations. However, these two factors are not
completely degenerate. Because by increasing the spot size the
finite size effect becomes important. This effect strongly alters
the shape of the spot-induced signals. Whereas, increasing the spot
temperature contrast only increases the perturbation signals and
does not change the shape of the signals. Hence, these two effect
are partially degenerate in the photometric measurements. Although,
two degenerate spots with the same photometric deviations have
different, intrinsic polarization signals, but by considering two
different magnetic fields, they can have the same polarization
signal and be polarimetrically degenerate. In Figure \ref{fig2c}, we
make two microlensing events with spotted sources in which the spots
have the same positions but different sizes, temperature differences
and magnetic fields. These spots have the same, intrinsic
polarization signals. These two events have the same light and
polarimetric curves and are degenerate. However, in the photometric
curve degeneracy is between the spot size and its temperature
contrast. Therefore, we can not uniquely extract the spot properties
from the photometric or polarimetric deviations.

(d) We investigate the finite size effect on the photometric and
polarimetric signals due to the spot. The spot-induced perturbations
decrease and lengthen with increasing the source radius. Because, by
integrating over the source surface, the sharp signals due to the
points located at the caustic line shrink. In Figure \ref{fig2d}, we
represent a microlensing event with three different amounts of
$\rho_{\star}$ and considering $r_{s}=0.3~R_{\star}$. The residual
for each source size is calculated from the spotless source of that
particular size.

\subsection{Perturbation on the angle of polarization}
\begin{figure}
\begin{center}
\psfig{file=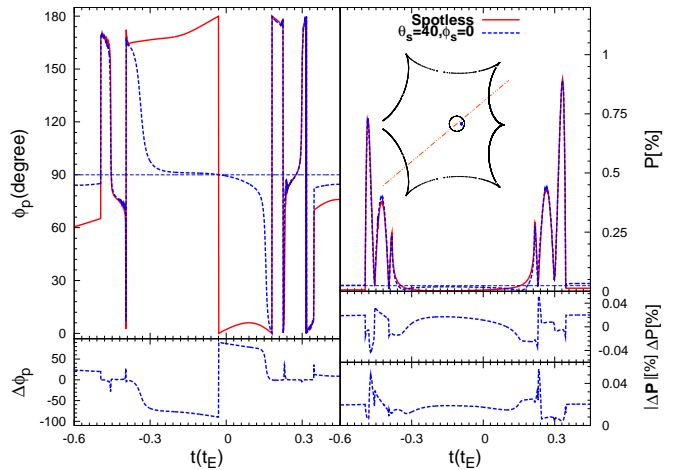,angle=270,width=9.cm,clip=} \caption{A
polarimetric microlensing event due a spotted source: The curves of
polarization angle $\phi_{p}$ and the polarization degree are shown
in the left and right panels. The simple models without spot effect
are shown by red solid lines. The polarimetric residuals with
respect to the simple models are plotted in the bottom of each
panel. The parameters used to make this microlensing event can be
found in Table (\ref{tab1}).} \label{fig3}
\end{center}
\end{figure}

The polarization angle $\phi_{p}$ strongly changes due to the
stellar spot. Because, this polarization angle is not related to the
total Stokes parameter, i.e. $S'_{I}$ which is very larger than the
polarized Stokes parameters. On the other hand, the polarization
angle $\theta_{p}$ alters a bit. Because, the broad band circular
polarization of spots is, one order of magnitude, smaller than the
linear ones. The spot-induced perturbation in the polarization angle
causes the polarization vector alters as follows:
\begin{eqnarray}\label{delp}
|\boldsymbol{\Delta P}|=[P'^2+P^2-2 P'P
\cos\theta_{p}'\cos(\phi_{p}'-\phi_{p})]^{1/2}.
\end{eqnarray}
Indeed, we define the polarization vector as $\boldsymbol{P}=(P
\cos\theta_{p}\cos\phi_{p},P\cos\theta_{p}\sin\phi_{p},P\sin\theta_{p})$.
Note that in normal microlensing events without spot effect
$\theta_{p}=0$. In Figure (\ref{fig3}) we show the curves of the
polarization angle $\phi_{p}$ (left panel) and polarization degree
(right panel) of a microlensing event.

Measuring the polarization angles helps us to indicate the different
components of the polarized intensity which gives some information
about the projection angles $\theta_{s}$ and $\phi_{s}$. On the
other hand, measuring the polarization angles in addition to the
polarization degree can increase detectability of stellar spots
through polarimetric measurements, e.g. in Figure \ref{fig2b}
although the maximum amount of the residual in the polarization
degree $\Delta P$ due to the spot reaches to $0.04$ per cent, but by
considering the variation of the polarization angles, the residual
in the polarization vector rises up to $0.35$ per cent.

\section{Observational prospects}\label{mont}
In the previous section we studied the spot-induced perturbations
during caustic-crossing in binary microlensing events. Now, we
estimate the probability of detecting the source spots in
polarimetric or photometric observations during caustic crossing. In
this regard, we perform a Monte Carlo simulation by generating many
synthetics caustic-crossing microlensing events with spotted and
giant sources. By considering two useful criteria for polarimetric
and photometric observations, we survey if the spot-induced
perturbations can be discerned in each light curve or polarimetric
curve. Finally, we estimate the number of events with detectable
spot-induced signals for given amounts of background stars and
observational time.

Our criterion for detectability of source spots in the microlensing
light curves is $(A'-A)/A\geq2$ per cent. For polarimetric
measurements, we assume that these observations are done by the
FOcal Reducer and low dispersion Spectrograph (FORS2) polarimeter at
Very Large Telescope (VLT) telescope. This polarimeter can achieve
to the polarimetric precision about $0.1$ per cent with one hour
exposure time from a source star brighter than $14.5$ mag
\cite{Ingrosso2015}. To evaluate the detectability of the
polarimetric signals due to the source spots, for each microlensing
event we evaluate signal to noise ratio (SNR) when the spot is
crossing the caustic curve, which is given by:
\begin{eqnarray}
SNR=\sqrt{[10^{-0.4
m_{\star}}A'+\Omega_{PSF}10^{-0.4\mu_{sky}}]t_{exp}10^{0.4m_{zp}}},
\end{eqnarray}
where, $m_{\star}(mag)$ is the apparent magnitude of the source
star, $A'$ is the magnification factor of the spotted source,
$\Omega_{PSF}$ is the area of the Point Spread Function (PSF),
$\mu_{sky}=22.5 (mag/arcs^{2})$ is the sky brightness, $t_{exp}(s)$
is the exposure time and $m_{zp}=28.3 (mag)$ is the zero point
magnitude of FORS2 polarimeter at VLT. We assume that there is no
blending effect. For detecting the spot signal, the exposure time
should be less than the time when the spot is crossing the caustic
curve and we set $t_{exp}=t_{E}\rho_{s}/3$ where $\rho_{s}$ is the
spot radius projected in the lens plane and normalized to the
Einstein radius. The threshold amount of SNR to achieve the
polarimetric precision about $0.1$ per cent is about
$SNR_{thr}=34000$. Hence, our criteria for detectability of the
source spot in the polarimetric curves are $|\boldsymbol{\Delta
P}|>0.1$ per cent as well as $SNR>SNR_{thr}$.

Here, the distribution functions of the binary lenses, source and
its spot parameters used to generate syntectic binary microlensing
events are illustrated. We explained the distribution functions for
the mass of the lenses, the Galactic coordinates and velocities of
both sources and lenses, the mass ratio for the binary lenses and
their distance as well as the source trajectory with respect to the
binary axis in our previous works \cite{sajadian12,sajadian15}, and
do not repeat them here.

We consider only giant stars as sources. For indicating the mass,
surface temperature, absolute magnitude and radius of giant sources,
we make an ensemble of giant stars using Besan\c{c}on model
\cite{besancon}. Then, we choose the source stars from that ensemble
randomly. We obtain the apparent magnitude of the source star
$(m_{\star})$ according to the absolute magnitude, distance modulus
and extinction. In that case, the more details can be found in
\cite{sajadian15}. Also, I consider constant amounts for
limb-darkening coefficients of source stars as $c_{1}=0.7$ and
$c_{2}=0.04$.

For the spot, we indicate its magnetic field and temperature
contrast using the equation (\ref{BT}) and adding Gaussian
fluctuations. There is a correlation relation between the magnetic
field of spots and the filling factor, i.e. the total fractional
area covered by the spot, as $f=0.4~B(kG)-0.7$  \cite{starspot}. In
the simulation we consider only one spot over the source surface.
For the Galactic bulge stars, only the biggest spot can probably be
detected. We need the distribution function of the radius of the
biggest spots over source stars. I assume that the ratio of the area
of the biggest spot to the total area due to all spots over the Sun
is the same as that ratio due to the spots over other stars. This
ratio for the Sun spots is about $0.04$ \cite{Solanki2003}.
Therefore, I first indicate the total fractional area due to the
spots, i.e. $f$, according to the magnetic field by adding a
Gaussian fluctuation and then multiply it by $0.04$ to indicate the
fractional area due to the biggest spot.
\begin{figure}
\begin{center}
\psfig{file=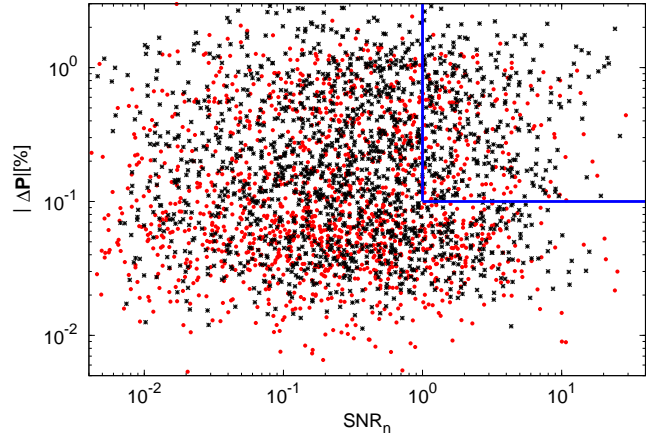,angle=0,width=9.cm,clip=} \caption{The scatter
plot of the spot-induced polarimetric signals in synthetic
microlensing events versus SNR which is normalized to the threshold
amount. The blue solid line separates the microlensing events with
detectable polarimetric signatures due to the spot. The black stars
represent the microlensing events with detectable photometric
signals due to the spot and red circles show the events in which the
spot-induced signatures can not be discerned through photometric
observations.} \label{fignew}
\end{center}
\end{figure}

In order to obtain the spot position in the lens plane, we need two
rotation angles of $\theta_{s}$ around $y$-axis and $\phi_{s}$
around $z$-axis. $\phi_{s}$ is uniformly taken in the range of
$[0,\pi]$. We choose $\theta_{s}$ by fixing the angular surface
element. Hence, we uniformly select $\psi$ in the range of $[-1,1]$
where $\theta_{s}=cos^{-1}\psi$. For the broadband Stokes
intensities due to the magnetic field, we first calculate them in a
spectral line and then by considering the factor $a$ (introduced in
the Appendix section) estimate the amounts of these Stokes
intensities. Finally, we investigate with how many probability
lensing magnifies these polarization signals and makes them be
observed.

Having performing the Monte Carlo simulation, we concluded that in
about $11.2$ per cent of the simulated events the spot-induced
polarimetric signals are detectable according to our criteria. Note
that the real efficiency is certainly less than this number. Indeed,
in observations of microlensing events, 3-5 consecutive data points
with deviations more than 2-3$\sigma$ with respect to the simple
model are needed to confirm a perturbation effect, where $\sigma$ is
the observational accuracy (see e.g. Dominik et al. 2010). But in
this Monte Carlo simulation, we do not simulate the data points.
Hence, short-duration spot-induced signals (e.g. see Figure
\ref{fig2c}) which are detectable according to our criteria would be
difficult to detect in real observations due to data spacing and may
be confused with the noise. On the other hand, observers should fit
many synthetic light and polarimetric curves with different
perturbations to the data points. Therefore during observation the
true event can be confused with other microlensing events which are
well fitted to the data points e.g. the perturbation due to the
stellar spots can be mistaken with a transit exo-planet around the
source star.

In Figure (\ref{fignew}), we represent the scatter plot of the
spot-induced polarimetric signals, i.e. $|\boldsymbol{\Delta P}|$
for synthetic microlensing events versus SNR normalized to the
threshold amount i.e. $SNR_{n}$. The blue solid line separates the
microlensing events with detectable polarimetric signatures due to
the spot from the others.

According to the OGLE-III data \cite{OGLE3}, about $f_{1}=18.4$ per
cent of the microlensing sources are giant stars and about
$f_{2}=5.4$ per cent of microlensing events are due to the binary
lenses with caustic-crossing features. Generally, all giant stars
are not magnetically active. RS CVn-type binaries, FK Comae-type
giants and Lithium rich giants have high magnetic activity
\cite{Korhonen2008}. Recently for some of other giant stars magnetic
fields have been detected and even mapped (e.g. Strassmeier et al.
1990, Auri\`{e}re et al. 2008, Dorch 2004). Indeed, the magnetic
activity decreases with the age \cite{Skumanich72}. Consequently, I
assume that about $f_{3}=1$ per cent of giant stars have stellar
spots and magnetic field stronger than $100$ G. The probability of
detecting a spot over a giant source through polarimetric
measurements in caustic-crossing of microlensing events can be given
by:
\begin{equation}
\tau_{s} = \epsilon ~ f_{1} ~ f_{2} ~ f_{3} ~ \tau,
\end{equation}
where $\epsilon=0.112$ is the polarimetric efficiency for detecting
spot-induced signals in the simulated events, $\tau$ is the optical
depth for the microlensing events towards a given direction. For the
direction of the Galactic Bulge $\tau=4.48\times 10^{-6}$
\cite{sum05}, hence the optical depth for the polarimetric spot
detection is $\tau_s=4.99\times 10^{-11}$. Finally, we estimate the
number of events for $N_{bg}$ background stars being detected during
the observational time $T_{obs}$ as
\begin{equation}
N_s = \frac{\pi}{2} \frac{T_{obs} N_{bg}}{<t_E>}\times \tau_s,
\end{equation}
where the average Einstein crossing time for the Galactic Bulge
events due to the giant sources is about $<t_E> = 24.6$~ days
\cite{OGLE3}. By monitoring $150$ million objects during $10$ years
towards the Galactic bugle, the number of binary microlensing events
with spotted and giant sources, which have detectable spot-induced
polarimetric signals, is $1.75$. Finally, we investigate in how
fraction of these simulated events the photometric spot-induced
signal $(A'-A)/A$ is greater than $2$ per cent. In this regard, the
photometric efficiency for detecting signatures due to the spot is
$52.6$ per cent. In that case the number of events during $10$ years
observations from $150$ million objects will be $8.19$. Note that,
we consider only giant stars as sources and do not consider
main-sequence stars. In Figure (\ref{fignew}) the black stars show
these events which have detectable photometric signatures due to the
spot.

Consequently, photometry observation is more efficient in detecting
stellar spots than the polarimetry observation. However, polarimetry
observations during microlensing events give us some information
about the magnetic field of the spots over the sources whereas the
photometric signals due to the spots do not depend on it. On the
other hand, the stellar spots change polarization orientation in
addition to the polarization degree whereas in the photometric
observations of spotted sources only the magnification factor
alters. Variation of the polarization angles give us extra
information about the different components of the polarized
intensity and as a result the position of spot on the source
surface. Also, the polarimetry observations can be complementary to
the photometry observations for detecting stellar spots with larger
amounts of the projection angle $\theta_{s}$.

\section{Conclusions}
Gravitational microlensing with finite size effect provides a
technique to probe the anomalies over the source surface specially
in caustic crossing. One of these anomalies is the stellar spot
which causes the light intensity of source star does not obey from
the limb-darkening function. It creates some perturbations in the
microlensing light curves.

The stellar spots also create a net polarization for the source
star. Although most spot-induced polarimetric signals of Galactic
bulge sources are too weak to be detected, but the lensing effect
can magnify the spot-induced polarization signal and makes it be
observed. However, lensing effect itself generates a net
polarization for the source star. Here, we investigated the
possibility of detecting the source spot through polarimetric
observations during caustic-crossing in microlensing events. In this
regard, we studied how the stellar spot perturbs polarimetric and
photometric curves in binary microlensing events during
caustic-crossing.

The stellar spot perturbs the microlensing light curves by making
positive and negative deviations \cite{Han2010}. If the time when
the photometric deviation due to spot is zero (between positive and
negative deviations) is inferred from microlensing light curves, we
can indicate the magnification factor of the spot which helps us to
characterize the spot properties except the temperature contrast.

We figured out some points about the spot-induced perturbations on
the polarization degree explained in following:

(a) The cooler spots have the higher magnetic fields and larger
polarization signals, but they have the smaller contributions in the
total flux. We showed that these two properties of spots,
temperature contrast and magnetic field, have the same contributions
in the polarimetric signal so that increasing both together does not
change the polarization signal.

(b) The maximum deviation in the polarimetry curve due to the spot
happens when the spot is located at the source edge and source is
entering to the caustic curve from that edge (near the spot).

(c) There is a (partial) degeneracy for indicating the spot size,
its temperature contrast with respect to the host star and its
magnetic induction from the deviations in light or polarimetric
curves.

(d) The spot-induced deviations decrease and lengthen by enhancing
the finite size effect.

(e) The stellar spots alter the polarization degree as well as
strongly change its orientation. Variation of the second gives some
information about the spot projection angles.

Finally, we estimate the number of the source spots that can be
observed with this method towards the Galactic bulge. During $10$
years monitoring of $150$ million stars towards the Galactic bulge,
we can specify in the order of $2$ stellar spot over giant sources
with this method whereas about $8$ stellar spots over giants can be
confirmed through photometric observations. Hence, photometry
observation is more efficient in detecting stellar spots than the
polarimetry observation. However, polarimetry observations during
microlensing events give us some information about the magnetic
field of the spots over the sources whereas the photometric signals
due to the spots do not depend on it.

\textbf{Acknowledgment} I gratefully acknowledge the referee, David
Heyrovsk\'y, for noticing some important points and useful comments.
I would like to thank Sohrab Rahvar for his helpful comments and
careful reading the manuscript and Andres Asensio Ramos and Heidi
Korhonen for useful discussions and comments.



\appendix
\section{The broadband Stokes intensities due to the magnetic field of spot}\label{Append}
The aim of this appendix is to estimate the amount of polarized
Stokes intensities of spots with the magnetic field and take into
account this intensities in the polarimetric microlensing
calculations. A stellar spot produces a polarization signal through
the Zeeman effect due to its magnetic field on the spectral lines.
The polarized radiation can be obtained by solving the Stokes
transfer equation:
\begin{eqnarray}
-\frac{d\boldsymbol{I}}{d\tau_{c}}=\boldsymbol{K}\boldsymbol{I}-\boldsymbol{J},
\end{eqnarray}
where $\tau_{c}$ is line of sight continuum optical depth,
$\boldsymbol{I}$ is the Stokes vector, $\boldsymbol{K}$ is the total
opacity matrix and is given by:
\begin{eqnarray}
\boldsymbol{K}=\boldsymbol{1}+\eta_{0} \boldsymbol{\Phi},
\end{eqnarray}
and $\boldsymbol{J}$ is the total emission vector,
\begin{eqnarray}
\boldsymbol{J}=(S_{c}+\eta_{0}S_{l}\boldsymbol{\Phi})\boldsymbol{e}_{0},
\end{eqnarray}
where $\boldsymbol{1}$ is the unit matrix,  $S_{c}$ is the source
function in the unpolarized continuum spectra and
$\boldsymbol{e}_{0}=(1,0,0,0)^{\dagger}$. We take the
Minle-Eddington model for the stellar atmosphere where the line
source function $S_{l}=B_{0}+B_{1}\tau_{c}$ and $B_{0}+B_{1}=I_{c}$
and $I_{c}$ is the continuum intensity close a spectral line. Here,
$\eta_{0}$ is the ratio between a spectral line and continuous
absorption coefficients. $\boldsymbol{\Phi}$ is the line absorption
matrix which is given by \cite{Rees89}:
\begin{eqnarray}
\boldsymbol{\Phi}=\left(\begin{array}{c}\eta_{I}~~~~~~\eta_{Q}~~~~~\eta_{U}~~~~~\eta_{V}\\\eta_{Q}~~~~~\eta_{I}~~~~\rho_{V}~~~~-\rho_{U}\\
\eta_{U}~~~~~-\rho_{V}~~~~\eta_{I}~~~~\rho_{Q}\\
\eta_{V}~~~~~\rho_{U}~~~~-\rho_{Q}~~~~\eta_{I}\end{array}\right),
\end{eqnarray}
where
\begin{eqnarray}\label{A5}
\eta_{I}&=&\frac{1}{2}[\eta_{p}\sin^2\theta_{s}+\frac{1}{2}(\eta_{r}+\eta_{b})(1+\cos^{2}\theta_{s})],\nonumber\\
\eta_{Q}&=&\frac{1}{2}[\eta_{p}-\frac{1}{2}(\eta_{r}+\eta_{b})]\sin^{2}\theta_{s}\cos
2\phi_{s},\nonumber\\
\eta_{U}&=&\frac{1}{2}[\eta_{p}-\frac{1}{2}(\eta_{r}+\eta_{b})]\sin^{2}\theta_{s}\sin
2\phi_{s},\nonumber\\
\eta_{V}&=&\frac{1}{2}(\eta_{r}-\eta_{b})\cos\theta_{s},\nonumber\\
\rho_{Q}&=&\frac{1}{2}[\rho_{p}-\frac{1}{2}(\rho_{r}+\rho_{b})]\sin^{2}\theta_{s}\cos
2\phi_{s},\nonumber\\
\rho_{U}&=&\frac{1}{2}[\rho_{p}-\frac{1}{2}(\rho_{r}+\rho_{b})]\sin^{2}\theta_{s}\sin
2\phi_{s},\nonumber\\
\rho_{V}&=&\frac{1}{2}(\rho_{r}-\rho_{b})\cos\theta_{s},
\end{eqnarray}
where $\theta_{s}$ and $\phi_{s}$ represent the position of spot on
the surface of source. Here the absorption $\eta_{p,r,b}$ and
anomalous $\rho_{p,r,b}$ dispersion profiles are given by:
\begin{eqnarray}
\eta_{p}&=&\eta_{0}H(a,v),\nonumber\\
\eta_{b,r}&=&\eta_{0}H(a,v\pm v_{H}),\nonumber\\
\rho_{p}&=&2\eta_{0}F(a,v),\nonumber\\
\rho_{b,r}&=&2\eta_{0}F(a,v\pm v_{H}),
\end{eqnarray}
in which $a$ is the damping constant, $H(a,v)$ and $F(a,v)$ are
Voigt and Faradey-Voigt profiles.
$v=(\lambda-\lambda_{0})/\Delta\lambda_{D}$ and
$v_{H}=\Delta\lambda_{Z}/\Delta\lambda_{D}$ are the wavelength
separation from the spectral line center (i.e. $\lambda_{0}$) and
the wavelength shift due to the Zeeman effect. Both parameters are
normalized to the Doppler broadening (i.e. $\Delta\lambda_{D}$). The
second one is given by:
\begin{eqnarray}
\Delta\lambda_{Z}= \frac{e\lambda^{2}B~g_{eff}}{4\pi~m~c},
\end{eqnarray}
in which $B$ is the value of the magnetic field, $g_{eff}$ is the
effective Land\'{e} factor and $e$, $m$ and $c$ have their usual
meanings. In this work we used the averaged values of these
parameters: $a=0.1$, $\Delta\lambda_{D}=300m{\AA}$, $\eta_{0}=10$,
$\lambda=6000{\AA}$, $g_{eff}=1.5$ and $B_{0}=B_{1}=0.5~I_{c}$
\cite{Huovelin}.

By assuming some limitations for the stellar atmosphere, there are
analytical solutions for the Stokes intensities
\cite{Unno56,Auer77,Landolfi82}. By considering the following
limitations: (i) the magnetic field is constant in amount and
direction over the spot, (ii) linear dependence of the source
function with optical depth (i.e. $B_{0}$ and $B_{1}$) and (iii)
constant ratio of the line and continuous absorbtion coefficients,
the Stokes intensities are given by \cite{Landolfi82}:
\begin{eqnarray}\label{A7}
I_{\lambda}&=&B_{0}+\mu B_{1}\Delta^{-1}[(1+\eta_{I})((1+\eta_{I})^2+\rho_{Q}^2+\rho_{U}^2+\rho_{V}^2)],\nonumber\\
Q_{\lambda}&=&-\mu~B_{1}\Delta^{-1}[(1+\eta_{I})^{2}\eta_{Q}+(1+\eta_{I})(\eta_{V}\rho_{U}-\eta_{U}\rho_{V})+\nonumber\\
&+&\rho_{Q}(\eta_{Q}\rho_{Q}+\eta_{U}\rho_{U}+\eta_{V}\rho_{V})],\nonumber\\
U_{\lambda}&=&-\mu~B_{1}\Delta^{-1}[(1+\eta_{I})^{2}\eta_{U}+(1+\eta_{I})(\eta_{Q}\rho_{V}-\eta_{V}\rho_{Q})+\nonumber\\
&+&\rho_{U}(\eta_{Q}\rho_{Q}+\eta_{U}\rho_{U}+\eta_{V}\rho_{V})],\nonumber\\
V_{\lambda}&=&-\mu~B_{1}\Delta^{-1}[(1+\eta_{I})^{2}\eta_{V}+\nonumber\\&+&\rho_{V}(\eta_{Q}\rho_{Q}+\eta_{U}\rho_{U}+\eta_{V}\rho_{V})],
\end{eqnarray}
where the index of $\lambda$ refers to the Stokes intensities in a
spectral line and
\begin{eqnarray}
\Delta&=&(1+\eta_{I})^2[(1+\eta_{I})^2-\eta_{Q}^2-\eta_{U}^2-
\eta_{V}^2+\rho_{Q}^2+\nonumber\\&+&\rho_{U}^2+\rho_{V}^2]-[\eta_{Q}\rho_{Q}+\eta_{U}\rho_{U}+\eta_{V}\rho_{V}]^{2}.
\end{eqnarray}

The microlensing observations are done in optical band and we should
estimate the contribution of the spectral lines in the broadband
photometric mode observation. In this regard, we assume that the
broad band polarization can be approximated by using the
polarization in an average line profile. Then, we scale the resulted
Stokes intensity with a factor to obtain the Stokes intensities in a
given pass band. This method was first used to calculate the broad
band linear polarization in cool stars, (e.g. Calamai et al. (1975),
Huovelin \& Saar (1991), Saar \& Huovelin (1993)). Hence, we compare
the amounts of the spectral Stokes intensities with the amounts of
broadband for fixed amounts of the magnetic field and the projection
angle. Indeed, we assume the similar dependence of the broadband
Stokes intensities to angles of $\theta_{s}$ and $\phi_{s}$ while
keeping same magnetic field.

We represent the broadband Stokes intensities due to the spot
magnetic field (equation \ref{spott}) with $I_{i,s}=i_{\lambda} a$
(for $i=I,Q,U,V$), in which $a$ is the ratio of the broadband Stokes
intensities to the those in a spectral line.

Considering Milne-Eddington model for the source atmosphere, for the
case of $B=3.2~kG$ and $\theta=0^{\circ}$, the maximum amount of the
broadband circular polarization reaches to $0.1$ per cent
\cite{Stift96}, while the circular polarization in spectral line of
$6301 {\AA}$ is about $3-4$ per cent
\footnote{http://www.iac.es/proyecto/magnetism/pages/codes/milne-eddington-simulator.php}.
Therefore, the broadband circular polarization is one order of
magnitude less than the circular polarization in a spectral line.
Indeed, the Stokes V profile has approximately an anti-symmetric
shape versus $v$ in a spectral line. Hence, the net circular
polarization over a broad band would be negligible
\cite{Borrero2011}. For $B=4~kG$ and $\theta=60^{\circ}$, the
broadband linear and circular polarizations are about $0.8$ and
$0.02$ per cent \cite{Stift97}. For the spectral line of $6301
{\AA}$ they are almost $1$ and $0.7$ per cent. Comparing these
numbers, the broadband linear polarization is almost equals to the
linear polarization in the spectral line. Therefore, we set $a=1$
for linear Stokes intensities and $a=0.1$ for the circular Stokes
intensity.
\end{document}